\documentclass[aps,reprint,superscriptaddress,
nofootinbib,showpacs,amsmath,amssymb]{revtex4-1}

\usepackage{latexsym,graphicx,slashed,bm,dcolumn}
\usepackage{hyperref}

\newcommand\PM[1]{\begin{pmatrix}#1\end{pmatrix}}

\renewcommand{\to}{\rightarrow}

\def\ov{\overline}
\def\ee{\end{equation}}

\def\lsim{\raise0.3ex\hbox{$\;<$\kern-0.75em\raise-1.1ex
\hbox{$\sim\;$}}}
\def\gsim{\raise0.3ex\hbox{$\;>$\kern-0.75em\raise-1.1ex
\hbox{$\sim\;$}}}
\def\cal{\mathcal}

\def\cL{{\cal L}}

\def\cP{{\cal P}}
\newcommand{\vect}[1]{\mbox{\boldmath$#1$}}

\def\dm{ \epsilon_{n\overline{n}}  }   
\def\alp{\varepsilon_{nn'}}   
\def\bet{\varepsilon_{n\overline{n}'} }   
\newcommand{\nbar}{{\overline{n}} } 

\def\rB{{\rm B}}
\def\rL{{\rm L}}

\begin{document}

\title{ A possible shortcut for neutron--antineutron oscillation through mirror world}

\author{Zurab~Berezhiani}
\thanks{zurab.berezhiani@aquila.infn.it} 
\affiliation{Dipartimento di Fisica e Chimica, Universit\`a di L'Aquila, 67100 Coppito, L'Aquila, Italy} 
\affiliation{INFN, Laboratori Nazionali del Gran Sasso, 67010 Assergi,  L'Aquila, Italy}


\begin{abstract}
Existing bounds on the neutron-antineutron mass mixing, 
$\epsilon_{n\bar n} < {\rm few} \times 10^{-24}$~eV,  
impose a severe upper limit  on $n - \bar n$ transition probability, 
$P_{n\bar n}(t) < (t/0.1 ~{\rm s})^2 \times 10^{-18}$ or so,  where $t$ is the neutron flight time.    
Here we propose a new mechanism of $n- \bar n$ transition which is not induced by direct 
mass mixing $\epsilon_{n\bar n}$ ($\rB=2$) but is mediated instead by $\rB=1$   
mass mixings of  the neutron with the hypothetical states of mirror neutron 
$n'$ and mirror antineutron $\nbar'$.  The latter can be  as large as 
$\epsilon_{nn'}, \epsilon_{n\bar{n}'} \sim 10^{-15}$~eV or so, 
without contradicting present experimental limits and nuclear stability bounds. 
 The probabilities of $n-n'$ and $n-\bar{n}'$ transitions, $P_{nn'}$ and $P_{n\bar{n}'}$, 
depend on environmental conditions in mirror sector, and  they can be resonantly amplified 
by applying the magnetic field of the proper value. 
This opens up a possibility of $n-\bar n$ transition with the probability 
$P_{n\bar n} \simeq P_{nn'}  P_{n\bar{n}'}$ which can reach the values $\sim 10^{-8} $ or even larger.
For finding this effect  in real experiments, the magnetic field  should not be suppressed but properly varied. 
These mixings can be induced by new physics  at the scale of few TeV  which may also  
originate a new low scale co-baryogenesis mechanism between ordinary and mirror sectors. 
 
 \end{abstract}

\maketitle

\noindent
{\bf 1.} 
Discovery of neutron--antineutron ($n-\nbar$) oscillation \cite{Kuzmin,nnbar,Glashow} 
would be a clear evidence of the baryon number violation, and can shade more light 
on the origin of the matter--antimatter asymmetry in the Universe \cite{Sakh}. 
Nowadays this phenomenon is actively discussed (see \cite{Phillips,Babu1,Babu2} for reviews) 
and new projects   for its experimental  search are under consideration 
\cite{Milstead,Fomin,Nesvizhevsky,Gudkov,Addazi-ESS}. 

In the Standard Model  (SM) frames the neutron has only
the Dirac mass term $m \, \ov{n} n$ which conserves baryon number $\rB$.   
The oscillation $n-\nbar$ which violates $\rB$ by two units ($\Delta \rB =2$)
can emerge if the neutron has also Majorana mass term 
originated from new physics
\begin{equation}\label{dm}
\frac{\dm}{2} \big( n^T C n + \ov n\, C \ov{n}^T \big) 
= \frac{\dm}{2}  \big( \ov{n_c}\, n ~+~ {\rm h.c.}  \big) 
\end{equation} 
where $n$ and $n_c = C \ov{n}^T$ are the neutron and antineutron fields, 
$C$ being the charge conjugation matrix.   
The mass terms (\ref{dm}) can be induced via the effective six-fermion operators 
$\frac{1}{M^5}(udd)^2$ involving $u$ and $d$ quarks, with $M$ being a large cutoff scale 
of new physics.\footnote{ 
These operators can have different Lorentz and color structures and generically 
they induce  four bilinear terms
$\ov{n}n_c$, $\ov{n_c}n $,  $\ov{n}\gamma^5 n_c$, $\ov{n_c}\gamma^5 n$, 
with complex constants. However, by a proper redefinition of the 
fields, these terms can be reduced to just one combination (\ref{dm})
with a real $\dm$ which is explicitly invariant under C transformation $n\leftrightarrow n_c$ 
as well as under parity $n \to i \gamma^0 n$ \cite{Arkady}.}

Oscillation $n-\nbar$  is hampered by the medium effects as interactions 
with the residual gas and magnetic field. 
In particular, in the  presence of magnetic field  $B$ the oscillation probability 
for a free flight time $t$ is  \cite{Phillips}:\footnote{
In present experiments the free flight times are rather small, $t\sim 0.1$~s or so,  
and  the neutron decay can be neglected.} 
\begin{equation}\label{P-Omega}
P_{n\nbar}(t) =  \frac{\dm^2 \sin^2\big(\sqrt{\Omega^2 +\dm^2}\,t \big)}{\Omega^2 + \dm^2}   
\approx \frac{\ \sin^2\big(\Omega t \big)}{\tau_{n\nbar}^2\Omega^2}   
\end{equation} 
where $\Omega = \vert \mu B \vert = (B/0.5\,{\rm G})\times 4581~{\rm s}^{-1}$,    
with $\mu = 6.031\times 10^{-12}$~eV/G being the neutron magnetic moment,  
and $\tau_{n\nbar}=\dm^{-1}= (10^{-23}\, {\rm eV}/\dm) \times 0.66 \cdot 10^8$\,s  
is characteristic oscillation time. 
 The same formula  with $\Omega \to \vert U -\bar{U} \vert/2$ 
applies in the presence of matter,  $U$ and $\bar{U}$ being 
the matter induced optical potentials for $n$ and $\nbar$.  
If the medium effects are suppressed so that the
{\it quasi--free} condition $\Omega t \ll 1$ is fulfilled,  
then Eq. (\ref{P-Omega}) gives  
\begin{equation} \label{n-nbar} 
P_{n\nbar}(t) = (\dm\, t)^2 =  
\left(\frac{t}{0.1~{\rm s}}\right)^2 \left(\frac{10^8~{\rm s}}{\tau_{n\nbar}}\right)^2 \! \times 10^{-18} 
\end{equation} 
If medium effects are strong, $\Omega t \gg 1$, the probability  becomes smaller. 
The time dependent factor in (\ref{P-Omega}) can be averaged  
and we get: 
\begin{equation}\label{P-Omega-av}
\ov{P}_{n\nbar} = \frac{\dm^2}{2\Omega^2} = 
\left(\frac{0.5\, {\rm G}}{B}\right)^2  \left(\frac{10^8~{\rm s}}{\tau_{n\nbar}}\right)^2 \!
\times 2.4\cdot 10^{-24} 
\end{equation} 

Experiment on direct search for $n-\nbar$ oscillations was performed  
in Institut Laue-Langevin (ILL) using the cold neutron beam with intensity 
$10^{11}~n$/s propagating in a long vessel (80~m)  before reaching the 
antineutron detector, with a mean flight time $t \simeq 0.1$~s  \cite{Baldo}. 
The quasi-free condition $\Omega t \ll 1$ was fulfilled by suppressing 
 the magnetic field  ($B< 0.1$~mG) and the residual gas pressure 
($P<2\cdot 10^{-4}$~Pa) in the drift vessel. 
No antineutron was detected in a running time  of $2.4\times 10^7$~s, 
and the lower limit on $n-\nbar$ oscillation time 
$\tau_{n \nbar} > 0.86 \times 10^8$~s (90 \% C.L.) was reported    
which in turn translates into an upper limit on $n-\nbar$ mass mixing 
$\dm < 7.7 \times 10^{-24}$~eV. 

Somewhat stronger but indirect bounds are obtained  from the nuclear stability.  
E.g. experimental limits on oxygen decay yields 
$\tau_{n \nbar} > 2.7 \times 10^8$~s (90 \% C.L.) \cite{SK},   
or  $\dm < 2.5 \times 10^{-24}$~eV.  
Due to these bounds,  for realistic flight times $t \leq 0.1$
the probability of $n-\bar n$ conversion (\ref{n-nbar}) via the 
direct mixing (\ref{dm}) cannot exceed $10^{-18}$ or so. 

In this paper we show that effective channel for $n-\bar n$ conversion 
can be induced via $n \to n'/\nbar' \to \nbar$ transitions assuming that 
the neutron has $\Delta \rB=1$ mixings with the hypothetical   
neutron $n'$ and antineutron $\nbar'$ states belonging to dark mirror world. 
As it was shown in Refs. \cite{BB-nn',More}, $n-n'$ oscillation time $\tau_{nn'}$ can be  
smaller than 1~s so that $n-n'$  mixing mass can be as large as $\alp \sim 10^{-15}$~eV, 
or even larger. 
This possibility is not excluded neither by nuclear stability bounds nor by astrophysical 
and cosmological limits \cite{BB-nn'}, and it can have observable effects  for the 
propagation of the ultra-high energy cosmic rays at cosmological distances \cite{UHECR,Askhat}. 
This phenomenon may manifest in neutron disappearance ($n\to n'$) and regeneration 
($n \to n' \to n$) experiments perfectly accessible to present experimental capabilities
\cite{BB-nn',Pokot}. In fact, several dedicated experiments were performed 
\cite{Ban,Serebrov,Bodek,Serebrov2,Altarev,ILL,Abel}, 
and it is rather intriguing that some of them show anomalous deviations from null hypothesis 
indicating to $\tau_{nn'} \sim 20$~s or so \cite{ILL,Nesti}.  

In principle, the neutron can have a mass mixing with both $n'$ and $\nbar'$ states,  
  with comparable sizes,  $\bet \sim \alp$. 
Here we show that if both $\Delta \rB=1$ oscillations $n-n'$ and $n-\nbar'$  take place, 
with the respective probabilities $P_{nn'}$ and $P_{n\nbar'}$, then also $n-\nbar$ oscillation 
will be induced with the probability $P_{n\nbar} \simeq P_{nn'}  P_{n\nbar'}$. 
In `usual' experimental conditions the disappearance of free neutrons due to 
$n-n'$ and $n-\nbar'$ oscillations might skip the detection  
since the probabilities $P_{nn'}$ and $P_{n\nbar'}$  are suppressed by environmental effects  
(as e.g.  mirror magnetic field at the Earth \cite{More}), 
and hence also their doubly suppressed  product $P_{nn'}  P_{n\nbar'}$  
can be below the experimental bound on $n-\nbar$ transition  \cite{Baldo}.  
However, in the properly adjusted experimental conditions 
(e.g. the applied magnetic field)
both $P_{nn'}$ and $P_{n\nbar'}$ can be resonantly enhanced,      
and  their product  $P_{nn'}  P_{n\nbar'} $ 
can exceed by many orders of magnitude the benchmark value $P_{n\nbar} = 10^{-18}$ 
(\ref{n-nbar}) attainable to the direct $n-\nbar$ mixing (\ref{dm}).    

\medskip
\noindent 
{\bf 2.} 
There may exist a shadow sector of particles which is a mirror replica 
of our observable particle sector.   Then all known particles: quarks, leptons, Higgs etc. 
must have their mirror twins: quarks$'$, leptons$'$, Higgs$'$ etc. 
which are sterile with respect to the SM gauge interactions $SU(3)\times SU(2)\times U(1)$
but should possess  gauge interactions  of their own SM$'$ sector $SU(3)'\times SU(2)'\times U(1)'$  
(see e.g. reviews \cite{IJMPA,Alice,EPJST}, for a historical overview see \cite{Okun}). 
Such shadow matter, invisible in terms of ordinary photons but gravitationally coupled  to our matter, 
can be a part of cosmological dark matter  as its asymmetric and atomic (dissipative) fraction,    
with specific implications  for cosmological evolution, formation and structure of galaxies and stars, etc. 
\cite{Blin,Hodges,BDM,Acta,BCV,IV,BCV2,M-stars}.

More generically, one can consider the theory based on the product $G\times G'$ of two 
identical gauge factors (SM or some its extension),   
ordinary particles belonging to $G$ and mirror particles to $G'$.  
Total Lagrangian of two sectors can be presented as 
\begin{equation} \label{Lagr} 
\cL_{\rm tot} = \cL  + \cL' + \cL_{\rm mix} 
\end{equation} 
where $\cL$ and $\cL'$ describe two particle sectors whereas $\cL_{\rm mix}$ contains 
possible cross-interactions between ordinary and mirror particle species.  

The identical form of $\cL$ and $\cL'$ can be guaranteed by discrete $Z_2$ symmetry 
$G\leftrightarrow G'$ when all our particles (fermions, Higgses and gauge fields) 
exchange places with their mirror twins (`primed' fermions, Higgses and gauge fields).
If this symmetry is exact, then two sectors should have identical microphysics.
In particular, our particles and their mirror twins should have exactly equal masses. 
The fermions of two sectors can be exchanged {\it with} or {\it without} the chirality change. 
Here we shall concentrate on the former case, so called mirror parity $\cP$   
which exchanges our left-handed (LH)  fermions with their right-handed (RH) mirror twins. 
This symmetry can be interpreted as a generalization of parity.  

A  direct way to establish the existence of mirror matter is the 
experimental search for oscillation phenomena between ordinary and mirror particles.  
In fact, all neutral  particles, elementary (as neutrinos) or composite (as the neutron)  
can have mixings with their mirror twins.  
Such mixings violate lepton ($\rL,\rL'$) or baryon ($\rB,\rB'$) numbers of two sectors by one unit, 
and they can be originated from some relevant effective interactions
in the mixed Lagrangian $\cL_{\rm mix}$. 
Interestingly,  such interactions induce $\rB\!-\!\rL$ and $\rB'\!-\!\rL'$ violating reactions 
in the early universe which can co-generate  baryon asymmetries in both ordinary 
and mirror worlds \cite{BB-PRL,Bento:2002,Berezhiani-IJMP}.   

For warming up, let us consider  the lepton sector including three LH families 
of  our lepton doublets  $\ell =(\nu_L,e_L)^T$  and correspondingly three RH families of
mirror leptons $\ell'=(\nu'_R,e'_R)^T$ (family indices are suppressed). 
As it is well-known, effective D=5 operators $\frac{1}{M} \ell^T C \ell \phi \phi $ ($\Delta \rL=2$) 
induce the neutrino Majorana masses  $m_\nu \sim v^2/M$ \cite{Weinberg}, 
 where $M$ is some large cutoff scale, $\phi$ is the Higgs doublet of the SM 
 and $\langle \phi \rangle=v$ is its vacuum expectation value (VEV). 
 Then, by $\cP$ parity, mirror Lagrangian $\cL'$ should include 
 the analogous operators $\frac{1}{M} \ell^{\prime T} C \ell'  \phi' \phi'$ ($\Delta \rL'=2$) 
which induce the Majorana masses of mirror neutrinos $m_{\nu'} \sim v^{\prime2}/M$,  
 with $\phi$ being the Higgs doublet of SM$'$ and $\langle \phi' \rangle=v'$ being its VEV. 
 
However, also the mixed Lagrangian $\cL_{\rm mix}$ 
can include similar D=5 operators $\frac{1}{M} \ov{\ell} \ell' \phi^\dagger \phi'$  
which violate both $\rL$ and $\rL'$ by one unit (but conserve $\rL + \rL'$) 
and  induce $\nu-\nu'$ mass-mixings  $m_{\nu\nu'} \sim vv'/M$ \cite{ABS,FV,BM-nu}. 
Hence, mirror neutrinos can be natural candidates for sterile neutrinos,   
with specific oscillation pattern between three ordinary (active) and three mirror (sterile) neutrinos. 
In addition,  $\rL,\rL'$-violating scattering processes 
$\ell \phi \to \ell' \phi'$ etc. induced by these operators provide a 
co-leptogenesis mechanism \cite{BB-PRL,Bento:2002} 
 which can explain baryon and dark matter fractions in the universe, 
$\Omega'_{\rm B}/\Omega_{\rm B}\simeq 5$ \cite{Berezhiani-IJMP,NOW2012}.  

\medskip

\noindent
{\bf 3.} 
 Let us discuss now the system of ordinary and mirror neutrons.  
The neutron and antineutron Dirac masses are equal by fundamental reasons
(CPT):  $m \,\ov{n} n=m \,\ov{n_c} n_c$ and $m' \ov{n'} n'= m' \,\ov{n'_c} n'_c$.    
 As for ordinary and mirror neutron masses, they are equal     
 by mirror parity, $m'=m$, if the latter is an exact symmetry. 
 (In next section we consider also a situation when $n$ and $n'$ have a small mass splitting 
 due to spontaneous breaking of $\cP$ parity.)  
   
In principle, all four states, $n$, $n_c$, $n'$ and $n'_c$, can have mixings among each other.  
Namely, if one introduces the neutron--antineutron mixing (\ref{dm}),   
then, by mirror parity,  analogous mixing between the mirror neutron 
and mirror antineutron should also exist. 
Thus, the mass terms describing $\Delta\rB=2$ and $\Delta\rB'=2$ mixings read: 
  \begin{equation}\label{dm-mirror}
  \frac{\dm}{2}  \,\big(\ov{n_c}\, n + \ov{n'_c}\, n' \big)   ~+ {\rm h.c.}  
  \end{equation} 
As for mass mixings between ordinary and mirror states, 
they violate  both $\rB$ and $\rB'$ by one unit. 
However, $n-n'$ (and $n_c - n'_c$) mixing conserves the combination $\rB + \rB'$: 
\begin{equation}\label{alp}
 \frac{\alp}{2} \, \big(\ov{n'}  n   + \ov{n' _c}\, n_c  \big) ~ +  {\rm h.c.}
 \end{equation}
whereas $n-n'_c$ and $n'-n_c$ mixings conserve $\rB - \rB'$: 
\begin{equation}\label{bet}
\frac{\bet}{2}  \big( \ov{n'_c} \, n  +  \ov{n_c}\, n' \big)  ~ +  {\rm h.c.}  
 \end{equation} 
A priori, the mixing masses $\dm$, $\alp$ and $\bet$  are just independent phenomenological 
parameters which values can be limited only by the experiment or by astrophysical bounds. 
As already discussed above, the Majorana mass term (\ref{dm-mirror}) 
 is limited by the experimental bounds on nuclear stability  
as $\dm < 2.5 \times 10^{-24}$~eV \cite{SK}. 
As for the mixing terms (\ref{alp}) and (\ref{bet}),  they can be much larger: 
in fact, existing limits (discussed in more details in next sections)
do not exclude $\alp$ and $\bet$ as large as $10^{-15}$~eV or perhaps even larger. 
All these mixings can be induced by certain effective six-fermion operators, and 
to understand theoretically how they could naturally fall in the above ranges,   
one has to analyze how the structures of these operators  
are constrained by the gauge symmetry.  

In ordinary sector $SU(3)\times  SU(2)\times U(1)$ the quarks ($\rB =1/3$) 
have the LH components $q_L=(u_L,d_L)$ transforming as electroweak doublets 
and the RH ones $u_R,d_R$ transforming as singlets, 
whereas the anti-quark fields ($\rB =-1/3$) have opposite chiralities: 
$q^c_R=(u^c_R,d^c_R)$ and $u^c_L,d^c_L$. 
Mirror sector $SU(3)'\times  SU(2)'\times U(1)'$ 
must have exactly the same content  modulo the fermion chiralities: 
mirror quarks are  the RH doublets $q'_R=(u'_R,d'_R)$ and LH singlets $u'_L,d'_L$ ($\rB'=1/3$),  
and antiquarks are $q^{\prime c}_L = (u^{\prime c}_L,d^{\prime c}_L)$ 
and $u^{\prime c}_R,d^{\prime c}_R$ ($\rB'=-1/3$). 
 
Thus, gauge symmetries allow D=9 operators  with $\Delta \rB=2$ and  $\Delta \rB'=2$:  
 \begin{equation}\label{B2}
 \frac{1}{\Lambda_2^5} \left[ \overline{ (u^{c} d^{c} d^{c} )_L} (udd)_R + 
 \overline{  (u^{\prime c} d^{\prime c} d^{\prime c} )_R} (u'd'd')_L \right] \, +  {\rm h.c.} 
  \end{equation}
as well as the mixed D=9 operators conserving $\rB+\rB'$: 
\begin{equation} \label{B+B} 
\frac{1}{\Lambda^5} \, \overline{  (u'd' d')_L} \, ( u d d) _R   ~ + {\rm h.c.}  
\end{equation} 
where the parentheses contain gauge invariant spin 1/2 chiral combinations 
of three ordinary quarks which can be composed   as e.g. 
$(u_{R,L}^T Cd_{R,L}) d_R$, and analogous combinations  
$(u^{\prime T}_{L,R} C d'_{L,R}) d'_L$ of three mirror quarks. 
 
As for the mixed $\rB - \rB'$ conserving  operators, they cannot be induced without breaking 
electroweak symmetries in two sectors 
since $d_L\subset q_L$ and $d'_R\subset q'_R$ states reside in weak doublets. 
Thus, the gauge invariance of these operators requires insertion of the Higgs doublets, 
$\phi$ or $\phi'$,  and their minimal dimension is D=10. After inserting the VEVs 
$\langle \phi \rangle = \langle \phi' \rangle=v$, these operators read: 
 \begin{equation}\label{B-B} 
\frac{v}{\Lambda_1^6}    
\left[ \overline{  (u^{\prime c} d^{\prime c} d^{\prime c} )_L}  ( u d d)_R   + 
 \overline{ (u^{c} d^{c} d^{c} )_R} ( u' d' d') _L \right] +  {\rm h.c.} 
\end{equation} 
 
Operators (\ref{B2}),  (\ref{B+B}) and (\ref{B-B}) are explicitly invariant 
under $\cP$ parity $(udd)_{R.L} \leftrightarrow (u'd'd')_{L,R}$.\footnote{
The following remark is in order. The above operators are invariant under $\cP$  parity  
but not under usual P and C transformations,  and generically they induce 
also bilinear terms involving $\gamma^5$.  
However, the latter can be eliminated by proper redefinition of the fields 
as in Ref. \cite{Arkady}  and the mixing terms can be brought to the C-invariant 
forms (\ref{dm-mirror}), (\ref{alp}) and (\ref{bet}).  
The mixing masses  can generically be complex but  for simplicity we take them 
all real, postponing CP violating effects  to be discussed elsewhere.} 
  The cutoff scales $\Lambda$, $\Lambda_1$ and $\Lambda_2$ 
 are  a priori independent parameters which can be determined 
(and related to each other)   in the context of UV-complete renormalizable models 
employing some heavy intermediate particles.  

Namely, in the context of the see-saw type models for $n-n'$ mixing proposed in 
Refs. \cite{BB-nn',BM} one can introduce a gauge singlet Dirac Fermion $N$ 
which acts  as a messenger between two sectors, with the mass term 
 $M \ov{N}N = M \ov{N_L} N_R + {\rm h.c.}$, where $N_L$ and $N_R$ are 
 respectively its LH and RH components.    
Introducing also color-triplet (respectively of $SU(3)$ and $SU(3)'$) scalars $S$ and $S'$  
with masses $M_S=M_{S'}$ (equal due to $\cP$ parity), one can write 
the following Lagrangian terms: 
\begin{eqnarray}\label{Yuk} 
&& \cL_{\rm Yuk} = u_{R,L}^TCd_{R,L} S  + S^\dagger N_R^T Cd_R + {\rm h.c.}   
\nonumber \\ 
&&  \cL'_{\rm Yuk} = u^{\prime T}_{L,R}Cd'_{L,R} S' + S^{\prime\dagger} N_L^T C d'_L + {\rm h.c.}
\end{eqnarray} 
(the Yukawa constants and gauge indices are suppressed).
After integrating out the heavy states $S,S'$ and $N$, 
 these terms induce $\rB+\rB'$ conserving operator (\ref{B+B})  with  
 $\Lambda^5 \sim M_N M_S^4$ \cite{BB-nn',BM} which gives rise to mixing (\ref{alp}):  
\begin{equation}\label{alp1}
 \alp \sim \frac{\Lambda_{\rm QCD}^6}{\Lambda^5} \sim
 \frac{ (10~{\rm TeV})^5 }{MM_S^4} \times 10^{-15}~{\rm eV} 
\end{equation} 
 Operator  (\ref{B2}) can be induced in this model, with 
$\Lambda_2^5 \sim M^2 M_S^4/\mu$,  
if in addition to Dirac mass $M\ov{N}N$ the heavy fermion $N$  
has also a Majorana mass term $\mu N C N + {\rm h.c.}$  \cite{BB-nn',BM}. 
Provided that $\mu \ll M$, we get mixings (\ref{dm-mirror}) with  
\begin{equation}\label{dm1}
 \dm \sim \frac{\Lambda_{\rm QCD}^6}{\Lambda_2^5}\sim 
\frac{\mu \Lambda_{\rm QCD}^6}{M^2M_S^4}  \sim \frac{\mu}{M} \alp 
 \end{equation} 
Hence, one can naturally have $\dm \ll \alp$ 
if the Majorana mass $\mu$ is much smaller than the Dirac Mass $M$.
 In particular, small $\mu$ can be  induced by the low scale spontaneous baryon violation 
as e.g. in Refs. \cite{BM,B-L}. Clearly, $\dm$ vanishes if $\mu=0$. 

Finally, also  operator (\ref{B-B}) can be induced 
if we introduce an extra vector-like down-type quark $D$ 
having a Dirac mass term  $M_D (\ov{D_L}D_R  + {\rm h.c.})$,  
where both $LH$ and $RH$ components $D_L$ and $D_R$ are  
the weak singlets.\footnote{Extra vector-like species 
of these type are key ingredients for the fermion mass generation via Dirac seesaw 
mechanism \cite{PLB83,PLB85,Khlopov1,Khlopov2}.  
Recently it was pointed out  that the mixing  of $d$ quark 
with such $D$-type fermion with mass $\sim 1$~TeV 
can also help to explain the anomalies in the CKM matrix unitarity  \cite{CKM}. }  
Then, by $\cP$ parity, also its mirror twin $D'$ should be introduced 
with the same Dirac mass $M_D (\ov{D'_L}D'_R  + {\rm h.c.})$. 
Hence, the following Yukawa terms can be added to Lagrangian (\ref{Yuk}):
\begin{eqnarray}\label{Yuk-2} 
&& \Delta\cL_{\rm Yuk} = S\, N_R^T C D^c_R  + 
\phi D_R^T C q^c_R    +  {\rm h.c.}   \nonumber \\ 
&&  \Delta \cL'_{\rm Yuk} = 
S' N_L^T C D^{\prime c}_L  + \phi^{\prime} D^{\prime T}_L C q^{\prime c}_L + {\rm h.c.}
\end{eqnarray} 
In this way, after substituting the Higgs VEVs $\langle \phi \rangle = \langle \phi' \rangle =v$, 
operator (\ref{B-B}) is induced with $\Lambda_1^6 \sim M_D M M_S^4 \sim M_D\Lambda^5 $. 
So we get mixing term (\ref{bet}) with  
\begin{equation} \label{bet1}
\bet \sim \frac{v \Lambda_{\rm QCD}^6}{\Lambda_1^6} \sim \frac{v}{M_D} \alp
\end{equation} 
Due to the experimental bound on extra quark mass $M_D > 1$~TeV, one expects   
$\bet/\alp  < 0.1$ or so.  However, one cannot exclude a situation 
when $\alp$ and $\bet$  have comparable values, by some conspiracy of the Yukawa constants 
omitted in couplings (\ref{Yuk}) and (\ref{Yuk-2}). In the following, 
we conservatively  consider  the parameter $r=\bet/\alp$ to be less than 1   
and take $r=1$ (i.e. $\bet = \alp$) as a marginal case. 

Extra color particles, as scalar $S$ and fermion $D$ with the masses of few TeV,   
can be within the reach of the LHC or future accelerators.  
This makes our picture potentially  testable also at the high energy frontier.

\medskip
\noindent
{\bf 4.} 
In what follows we discuss the oscillation phenomena between the four states $n,\nbar, n', \nbar'$ 
due to mass mixings  (\ref{dm-mirror}), (\ref{alp}) and  (\ref{bet}). 
 The time evolution of the wavefunction $\Psi = (\psi_n, \psi_{\nbar }, \psi_{n'},\psi_{\nbar'})^T$
 in medium is determined  by the Schr\"odinger equation $i d\Psi/dt = H\Psi$ 
 with Hamiltonian\footnote{Here we assume 
 that the masses of ordinary and mirror neutrons are exactly equal, $m=m'$, 
 as well as their magnetic moments, $\mu=\mu'$,  
and omit the  identical mass terms in Hamiltonian (\ref{H44}) since they are not 
relevant for the oscillation processes.}  
\begin{equation}\label{H44}
 H=\PM{ U \! + \! \mu\vect{B}\vect{\sigma}   & \dm & \alp & \bet \\
\dm  & \! \bar{U}\!  - \! \mu\vect{B}\vect{\sigma}   & \bet  &  \alp \\
\alp &  \bet &  \! U' \! + \! \mu\vect{B}'\!\vect{\sigma}  & \dm  \\
\bet & \alp  & \dm  &  \! \bar{U}' \!  - \! \mu\vect{B}'\! \vect{\sigma} }    
\end{equation}
where each component is in itself a $2\times 2$ matrix 
acting on two spin states as far as wavefunctions 
$\psi_n,\psi_{\nbar}$ etc. are two component spinors.   
The terms $U,\bar{U},U'$ and $\bar{U}'$  describe 
the matter induced potentials respectively for $n,\nbar, n'$ and $\nbar'$ states,   
 $\vect{B}$ and $\vect{B}'$ are ordinary and mirror magnetic fields and  
$\vect{\sigma}=(\sigma_x,\sigma_y,\sigma_z)$ are the Pauli matrices.   
 Without loss of generality, direction of ordinary magnetic field can be taken  
as $z$-axis, i.e. $\vect{B}=B (0,0,1)$, 
and the mirror one can be taken in $x-z$ plane,   
i.e. $\vect{B}'=B'(\sin\!\beta,0,\cos\!\beta)$. 
 
The Hamiltonian eigenstates $n_{1,2,3,4}$ are superpositions of the `flavor' states $n,\nbar,n',\nbar'$: 
\begin{equation}\label{S}
\left(\begin{array}{c} n_1 \\ n_2 \\ n_3 \\ n_4 \end{array}\right) = 
\PM{ S_{1n}   &  S_{1\nbar}  &  S_{1n'}  &  S_{1\nbar'}  \\
S_{2n}   & S_{2\nbar}   & S_{2n'}  &  S_{2\nbar'}  \\
S_{3n}  & S_{3\nbar}  &   S_{3n'}  & S_{3\nbar'}  \\
S_{4n}  & S_{4\nbar }  &   S_{4n'}  & S_{4\nbar'}  } 
\left(\begin{array}{l} n \\ \nbar  \\  n' \\ \nbar' \end{array}\right)  
\end{equation} 
This mixing matrix determines the oscillation probabilities, 
as probability $P_{n\nbar}(t)=\vert \psi_{\nbar}(t) \vert^2$ 
to find  antineutron after time $t$ 
if the initial state at $t=0$ is the neutron, $\Psi(0) = (1,0,0,0)^T$, 
as well as the conversion probabilities $P_{nn'}(t)$ and $P_{n\nbar'}(t)$ 
 into mirror states  $n'$ and $\nbar'$. 

 The oscillations can be averaged in time if their periods are shorter than the flight time. 
The mean probabilities  can be readily deduced, considering that  
 creation of $n$ means to create the eigenstates $n_{i}$
with the respective probabilities $\vert S_{in}\vert^2$. 
The eigenstates do not oscillate between each other but just propagate, 
and an eigenstate $n_i$ is detectable as $\nbar$  with the probability $\vert S_{i\nbar}\vert^2$. 
Hence,  for mean probability of $n-\nbar$ conversion we get 
$\ov{P}_{n\nbar}\! = \!\sum_{i}\vert S_{in}\vert^2 \vert S_{i\nbar}\vert^2 $, 
and similarly for $\ov{P}_{nn'}$ and $\ov{P}_{n\nbar'}$. 

 In the rest of this section, we assume that the mirror gas effects 
at the Earth are negligible, $U',\bar{U}'=0$, and also suppose that 
the normal gas density is properly suppressed in laboratory conditions, $U,\bar U =0$.  
  So we leave only magnetic contributions in diagonal terms  of Hamiltonian (\ref{H44}), 
denoting  their moduli as $\Omega= \vert \mu B \vert$ and $\Omega' = \vert \mu B' \vert$.
 In addition, we assume that $\alp,\bet \gg \dm$ and neglect direct $n-\nbar$ mixing
 taking the limit $\dm=0$.
 
 Let us first discuss the situation under the minimal hypothesis adopted in Ref. \cite{BB-nn'}, 
assuming that  mirror magnetic field  $B'$ is negligibly small at the Earth and 
setting $\Omega'=0$.  
In this case the oscillations are suppressed by ordinary magnetic field $B$. 
However, by properly screening  the latter, the quasi-free condition $\Omega t \ll 1$ can be achieved  
 in which case the oscillation probabilities for a flight time $t$ 
become:\footnote{Let us remark that a complete formula for $P_{n\nbar}(t)$ with $\dm\neq 0$  
can be obtained simply by adding  to $P_{n\nbar}(t)$ in Eqs. (\ref{Probs-t}) 
the direct oscillation  probability $t^2/\tau_{n\nbar}^2$ (\ref{n-nbar}). }
\begin{eqnarray}\label{Probs-t} 
 && P_{nn'}(t) = (t/\tau_{nn'})^2, \quad  P_{n\nbar'}(t) = (t/\tau_{n\nbar'})^2,    \nonumber \\ 
&& P_{n\nbar}(t) = P_{nn'}(t) P_{n\nbar'}(t)  = ( t^2/\tau_{nn'}\tau_{n\nbar'})^2     
 \end{eqnarray} 
where $\tau_{nn'} = \alp^{-1}$ and $\tau_{n\nbar'} = \bet^{-1}$ are 
characteristic oscillation times. 
In this case the direct bound $P_{n\nbar}(t) < 10^{-18}$ from the ILL experiment \cite{Baldo} 
 (performed with $t\approx 0.1$~s 
 in conditions of screened magnetic field $B< 0.1$~mG, i.e. $\Omega t < 0.1$), 
merely implies a lower limit for the product of oscillation times: 
 \begin{equation}\label{B=0}
\tau_{nn'} \tau_{n\nbar'} > 10^7~{\rm s}^2      
\end{equation} 
In large magnetic field  transition probabilities will be suppressed. 
Namely, for  $\Omega t \gg 1$,  oscillations can be averaged and one gets  
$\ov{P}_{nn'}= 2\alp^2/\Omega^2$, $\ov{P}_{n\nbar'}= 2\bet^2/\Omega^2$ and 
$\ov{P}_{n\nbar} = \ov{P}_{nn'} \ov{P}_{n\nbar'}= 4\alp^2\bet^2/\Omega^4$.

Independent limits on oscillation times  can be  obtained from dedicated experiments \cite{Ban,Serebrov,Bodek,Serebrov2,Altarev,ILL,Abel} 
 searching for anomalous losses of ultra-cold neutrons (UCN) in different  magnetic fields. 
 In the presence of both $n-n'$ and $n-\nbar'$ transitions, the neutron disappearance
 probability $P_{nn'} + P_{n\bar n'}$ is proportional to the sum $\alp^2+\bet^2$.  
 Thus the limits of these experiments 
should be applied to the effective combination $\tau = (\alp^2+\bet^2)^{-1/2}$.  
 
Considering the ratio $r = \bet/\alp$ as a free parameter,  we have 
$\tau_{n\nbar'} =  \tau_{nn'}/r$, $\tau = \tau_{nn'}/\sqrt{1+r^2}$ and so 
 \begin{equation}\label{r}
 \tau_{nn'}  \tau_{n\nbar'}  =   (r + r^{-1})\, \tau^2   \geq  2 \tau^2 
\end{equation}
Hence,  for typical values $r \sim 0.1$ one gets  $\tau_{nn'} \approx \tau $ and 
$\tau_{nn'}  \tau_{n\nbar'} \sim 10\, \tau^2$.
The product (\ref{r}) becomes minimal for the marginal case $r=1$, 
i.e. $\bet = \alp$, when $\tau_{nn'} = \tau_{n\nbar'} =  \sqrt2 \tau$ and   
$\tau_{nn'}  \tau_{n\nbar'} =2 \tau^2$. 
 
Under the minimal hypothesis ($B'=0$), 
experiments  measuring the UCN losses in the conditions of small ($\Omega t \ll 1$) 
and large ($\Omega t \gg 1$) magnetic fields \cite{Ban,Serebrov,Bodek,Serebrov2,Altarev,ILL,Abel}  
give rather stringent limits  on the effective time $\tau$.  
Namely,   experiments \cite{Serebrov,Serebrov2} 
performed  at the ILL by Serebrov's group give a combined  lower bound   
$\tau  > 448$~s (90\% C.L.) \cite{Serebrov2}.\footnote{Comparable limit  
$\tau  > 352$~s (95\% C.L.) was obtained in a recent experiment  
 performed at the Paul Scherrer Institut (PSI)   \cite{Abel}. } 
Eq. (\ref{r}) turns this limit  into $r$-dependent limit 
$ \tau_{nn'}  \tau_{n\nbar'}  >  (r+ r^{-1}) \times 2 \cdot 10^5$~s$^2$, 
which becomes comparable with the lower limit (\ref{B=0}) for $r = 0.02$ or so. 

However, these bounds on $\tau$ as well as the limit (\ref{B=0})  
become invalid if mirror field $B'$ is non-zero \cite{More}.   
In difference from normal magnetic field,   it cannot be screened in experiments,
and its contribution would block the neutron oscillations into $n'$ and $\nbar'$
(and also induced $n-\nbar$ oscillation)   
even if ordinary field $B$ is suppressed.   
     
Hamiltonian (\ref{H44}),  with non-zero magnetic fields $\vect{B}$ and $\vect{B'}$ 
having arbitrary directions, can be diagonalized following the techniques described 
in Refs. \cite{More,Nesti} and the elements of mixing matrix (\ref{S}) can be calculated. 
Let us also remark that the oscillation probabilities do not depend on the neutron 
spin state (polarization). 
If the difference of magnetic fields is large enough, $\vert \Omega-\Omega'\vert t \gg 1$,   
oscillations can be averaged in time,   
and the mean probabilities  can be presented  in a convenient form as 
  \begin{eqnarray}\label{magn}
&& \ov{P}_{nn'} = \frac{2\cos^2(\beta/2)}{ \tau_{nn'}^2 (\Omega - \Omega^{\prime })^2} + 
\frac{2 \sin^2(\beta/2)}{\tau_{nn'}^2(\Omega + \Omega^{\prime })^2} \, , \nonumber \\
&& \ov{P}_{n\bar n'} = 
\frac{2 \sin^2(\beta/2)}{\tau_{n\nbar'}^2(\Omega - \Omega^{\prime })^2} + 
\frac{2 \cos^2(\beta/2)}{\tau_{n\nbar'}^2(\Omega + \Omega^{\prime })^2} \,,  \nonumber \\
&& \ov{P}_{n\bar n} =  \ov{P}_{nn'} \ov{P}_{n\bar n'} 
\end{eqnarray}
Namely, in the limit  $B = 0$ we get $\ov{P}_{nn'}= 2/(\tau_{nn'}\Omega')^2$, 
$\ov{P}_{n\bar n'} = 2/(\tau_{n\nbar'}\Omega')^2$ and 
$\ov{P}_{n\bar n} = 4/(\tau_{nn'}\tau_{n\nbar'} \Omega^{\prime2})^2$. 
 Hence, in this case the bound from $n-\nbar$  experiment  $\ov{P}_{n\bar n} < 10^{-18}$ \cite{Baldo} 
 restricts the product of oscillation times  as  
\begin{equation}\label{product}
\tau_{nn'} \, \tau_{n\nbar'} > \frac{2 \times 10^9}{\Omega^{\prime2}} \approx 
\left(\frac{0.5\,{\rm G}}{B'}\right)^2 \times 95~{\rm s}^2 
\end{equation}
Note that this limit  is much weaker than the limit (\ref{B=0})  
obtained under the minimal hypothesis of vanishing $B'$. 

By increasing ordinary field $B$, $\Omega-\Omega'$ dependent terms in 
 $P_{nn'}$ and $P_{n\nbar'}$ start to increase and approach the resonance 
 when $B$ gets close to $B'$. For $\vert B-B'\vert$ large enough, 
 so that  $\vert \Omega -\Omega' \vert t \gg 1$, the Eqs. (\ref{magn}) 
 for time averaged probabilities remain applicable.   
However,  when $B$ gets so close to $B'$ that the quasi-free condition 
$\vert \Omega-\Omega'\vert t \ll 1$ is satisfied (e.g. $\vert B- B' \vert < 0.1$~mG for $t=0.1$~s), 
oscillation probabilities become maximal. Namely,  neglecting contributions of $\Omega+\Omega'$ 
dependent terms in (\ref{magn}) we get:
\begin{eqnarray}\label{magn-res}
&& P_{nn'}(t) =  \frac{t^2 \cos^2(\beta/2)}{ \tau_{nn'}^2 }, \quad 
P_{n\nbar'}(t) =  \frac{t^2 \sin^2(\beta/2)}{ \tau_{n\nbar'}^2 } \, , \nonumber \\
&& P_{n\nbar}(t) = P_{nn'}(t) P_{n\nbar'}(t) = 
\left(\!\frac{t^2}{\tau_{nn'}\tau_{n\nbar'}}\! \right)^2 \frac{\sin^2\!\beta}{4} 
 \end{eqnarray}

For non-vanishing $B'$, the limits on the effective oscillation time $\tau$  
were obtained by experiments \cite{Serebrov2,Altarev,ILL,Abel} 
which measured the UCN loss rates for different values and orientations 
of magnetic field $B$. 
Since the neutron disappearance probability is contributed by both $n-n'$ and $n-\nbar'$ 
oscillations, these experiments measure the sum of probabilities 
$P_{nn'}+P_{n\nbar'}$. Interestingly, the dependence on the unknown angle $\beta$ 
can be excluded by measuring the UCN losses,  for a given value of $B$, 
 for two opposite directions of the magnetic field, $\vect{B}$ and $-\vect{B}$.  
The inversion of magnetic field $\vect{B}\to -\vect{B}$ is equivalent to 
$\beta \to 180^\circ\!-\!\beta$, which exchanges the factors  $\cos^2(\beta/2)$ and $\sin^2(\beta/2)$ 
between the first and second terms in (\ref{magn}). Hence, the average 
result between $\vect{B}$ and $-\vect{B}$ measurements should not depend on $\beta$. 
Namely, in the ``off-resonance" case $\vert \Omega-\Omega' \vert t \gg 1$,  
 for these averages we get: 
 \begin{equation}\label{magn-av}
\ov{P}^{\rm av}_{nn'} + \ov{P}^{\rm av}_{n\nbar'}=
 \frac{1}{ \tau^2} \left[\frac{1}{(\Omega - \Omega^{\prime })^2} + 
\frac{1 }{(\Omega + \Omega^{\prime })^2} \right]  
\end{equation} 
and in ``close to resonance" case $\vert \Omega-\Omega' \vert t \ll 1$ , 
from (\ref{magn-res}) one analogously gets 
$\ov{P}^{\rm av}_{nn'} + \ov{P}^{\rm av}_{n\nbar'}= \frac12(t/\tau)^2$. 

The lower limit on the effective oscillation time $\tau$ obtained from  the the UCN  experiments 
\cite{Ban,Serebrov,Bodek,Serebrov2,Altarev,ILL,Abel} depends on the inferred value of $B'$. 
The results of these experiments are summarized in Ref. \cite{ILL,Abel} where  
the respective limits on $\tau$ are shown (see e.g. Fig.~7 in Ref. \cite{ILL})
as wiggly functions of $B'$ with the values varying between $\tau_{B'}=(20\div 40)$~s 
within the interval $B'= (30 \div 250)$~mG, excluding narrow peaks with heights exceeding 100 s 
at $B'= 100$~mG and $B'=200$~mG. (Therefore, for certain values of $B'$ in this interval, 
the $r$-dependent limits $\tau_{nn'}\tau_{n\nbar'} > 400 (r+r^{-1}) \times (\tau_{B'}/20\,{\rm s})^2$ 
obtained from Eq. (\ref{r}) become  competitive to $n-\nbar$ limit (\ref{product}).) 
However,  already at $B'=0.3$~G, the lower limit becomes much smaller, $\tau\approx 5$~s, 
and then it drops rapidly as $\tau \approx (0.5\,{\rm G}/B')^2 \times 1$~s or so. 
 
Let us turn now to the induced $n-\nbar$ oscillation, 
and investigate how large it probability can be.  
As noted above, mirror magnetic field cannot be controlled in the experiments and its value 
is unknown.  However, the ordinary field can be varied,   
and when $B$ gets closer to $B'$ so that $\vert \Omega - \Omega' \vert $ becomes small, 
the  probabilities $P_{nn'}$ and $P_{n\nbar'}$  resonantly increase. 
Then the probability of  induced $n-\nbar$ transition $P_{n\nbar} = P_{nn'} P_{n\nbar'}$ 
increases with a double velocity and it can become very large. 

For demonstration, let us take e.g.  $B'=0.5$~G and  $\tau_{nn'} \, \tau_{n\nbar'}=100$~s$^2$,    
so that in very small magnetic field $B\ll B'$ the probability of $n-\nbar$ conversion 
matches the ILL experimental bound $\ov{P}_{n\nbar}(B\!=\!0)  \simeq  10^{-18}$ \cite{Baldo}. 
For simplicity, we also take $\beta=90^\circ$. 
Let us discuss now what happens when $B$ increases and gets closer to $B'$, 
e.g.  $B=0.45$~G (i.e. $\vert B - B' \vert = 50$~mG).  
 Then from Eq. (\ref{magn}) we get $\ov{P}_{n\nbar}(B) =  2.5 \times 10^{-15}$  
 which is more than 3 orders of magnitude larger than $\ov{P}_{n\nbar}(B\!=\!0)$. 
 Moving closer to the resonance and taking e.g. $\vert B - B' \vert = 10$~mG, 
one gets $\ov{P}_{n\nbar}(B) = 1.5 \times 10^{-12}$, thus gaining another 3 orders of magnitude. 
 
By fine scanning over the magnetic field values one can arrive enough 
close to the resonance, achieving the quasi-free regime $\vert \Omega-\Omega'\vert t \ll 1$ 
(e.g. $\vert B- B' \vert < 1$~mG for $t=0.1$~s).  
In this case $n-\nbar$ transition probability   given by Eq. (\ref{magn-res})  
 can be rendered as large as 
\begin{eqnarray}\label{n-nbar-new}
&& P_{n\nbar}(t)   = \frac{\sin^2\!\beta}{4}  \left(\frac{t}{0.1~{\rm s}}\right)^{\!4} 
\! \left(\frac{100~{\rm s}^2}{\tau_{nn'}  \tau_{n\nbar'} }\right)^{\!2}\! \times 10^{-8} 
\nonumber \\
&& \quad\quad \quad  \leq  \sin^2\!\beta \left(\frac{t}{0.1~{\rm s}}\right)^{\!4} 
\left(\frac{B'} {0.5\,{\rm G}}\right)^{\!4} \times 2.5\cdot 10^{-9} 
\end{eqnarray} 
where the last row takes in account the limit (\ref{product}) on $\tau_{nn'}  \tau_{n\nbar'}$.  
Hence, taking e.g. $B'=0.5$~G and $\sin^2\beta = 1/2$ as random average of unknown angle,
we see that, for the neutron flight time $t\approx 0.1$~s,  the probability of 
 $n-\nbar$ transition induced by $n-n'$ and $n-\nbar'$ mixings  
can be 9 orders of magnitude larger than 
 the benchmark probability (\ref{n-nbar}) due to direct $n-\nbar$ mixing.
The effect can be stronger if $B'$ is larger than 0.5~G. 
E.g. for $B'= 1.5$~G, one  could gain another two orders of magnitude. 

Therefore, if mirror magnetic field is present at the Earth, the existing experimental limits 
do not exclude the possibility of enormous enhancement  of $n-\nbar$ conversion probability, 
by many orders of magnitude. This effect can be searched in experiments similar 
to the ILL experiment \cite{Baldo} but without suppressing the magnetic field $B$.  
Since the value of $B'$ is unknown, 
the applied magnetic field should be varied in some proper steps to find the effect of resonant 
enhancement.  As for the its direction (unknown angle $\beta$), 
it can be identified e.g. by changing the direction of applied magnetic field   
by 90 degrees: if by chance $\sin^2\beta$ is small, then $\sin^2(\beta+90^\circ)\approx 1$.

 \begin{figure}[t]
 \includegraphics[scale=0.3]{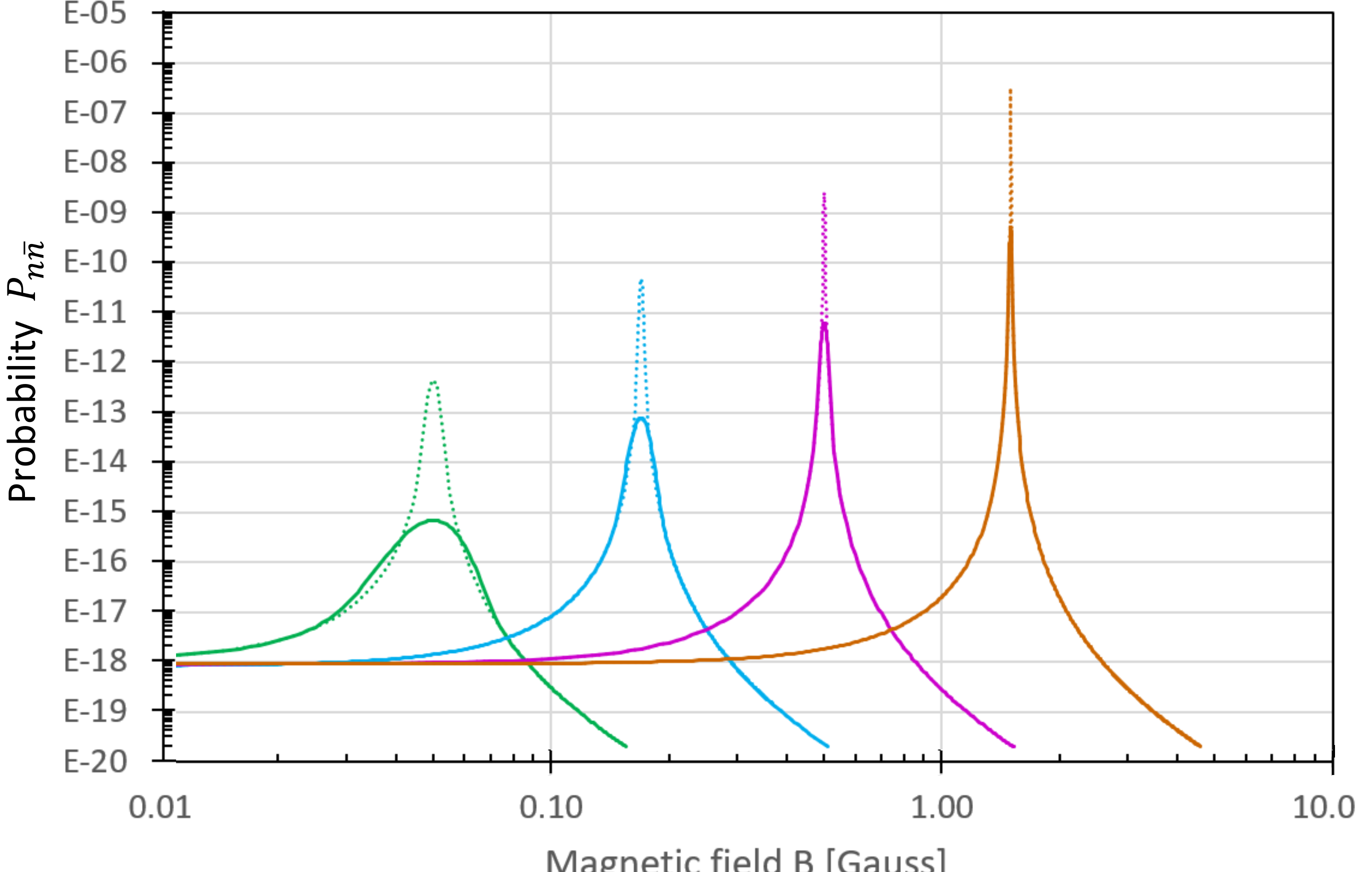}
 \caption{$P_{n\nbar}(t)$  as function of applied magnetic field $B$ 
for the cases $B'=0.05,\, 0.17,\, 0.5, \, 1.5$~G (from left to right). Dashed curves correspond 
to the flight time $t=0.1$~s, and solid curves correspond to shorter flight time $t=0.02$~s.  
For simplicity,  $\beta=90^\circ$ is assumed. }
\label{fig-Yura}
\end{figure}

In Fig. \ref{fig-Yura} we show how large effects can be obtained  
by scanning of the magnetic field $B$ in the cases of  different inferred values for mirror field $B'$.  
 For each choice, the product $\tau_{nn'}  \tau_{n\nbar'}$ is taken  to saturate the bound (\ref{product}),   
so that respective probabilities of $n-\nbar$ oscillation at vanishing magnetic field are normalized as 
 $P_{n\nbar}(B\!=\!0) \approx 10^{-18}$, fulfilling the bound from the ILL experiment \cite{Baldo}. 
The dotted curves of different colors show respective values of $P_{n\nbar}$ as functions of $B$ 
in the case of long baseline experiment with cold neutrons assuming a typical 
flight time $t=0.1$~s  (in some sense, repeating the ILL seminal experiment with about 100 m long 
vessel \cite{Baldo}  but with varying magnetic field). 
For comparison, we also show the similar expectations for a smaller scale experiment 
with shorter flight time $t=0.02$ (solid curves of the same colors).  
Taking in consideration that similar experiment  
 for the neutron regeneration effect $n\to n' \to n$ \cite{Frost}  
 is planned at the Oak Ridge National Laboratory (ORNL), with about 20 m baseline \cite{ORNL}, 
 the effect  $n\to n'/\nbar' \to \nbar$ regeneration into the antineutron 
can be searched along the same lines, just substituting the neutron detector by the antineutron one.  

We discussed the effect on $n-\nbar$ oscillations induced by $n-n'$ and $n-\nbar'$ mass mixings. 
Let us remark that such oscillations could occur also 
via the neutron transitional magnetic (or electric) dipole moments in 
$n'$ and $\nbar'$  states \cite{MDPI} which  possibility will be discussed in details 
elsewhere. Let us remind that the Lorentz invariance forbids the transitional moments 
between $n$ and $\nbar$ states but not between the ordinary and mirror states \cite{Arkady2}.  

Let us briefly discuss now the situation when ordinary and mirror neutrons 
are not exactly degenerate in mass but they a mass splitting 
$\delta  = m'-m \neq 0$ due to a spontaneous breaking of mirror parity 
which induces a difference between the Higgs VEVs 
$\langle \phi \rangle=v$ and $\langle \phi' \rangle=v'$. 
This possibility can related to asymmetric post-inflationary reheating of 
the ordinary and mirror sectors  \cite{BDM,Acta},  
and in some scenarios this splitting can be rather tiny \cite{Moh-Nus}.  

Interestingly,  $n-n'$  oscillation  with a mass splitting  
$\delta  \sim 10^{-7}\div 10^{-6}$~eV \cite{lifetime} 
can be a solution to the neutron lifetime puzzle related to about $4\sigma$ 
discrepancy between the neutron decay times measured in the trap and beam experiments.   
It is based on the idea that $n-n'$ conversion can be resonantly enhanced in strong (few Tesla) 
magnetic fields used in beam experiments and thus the deficit of protons is because  
mirror neutrons decay in invisible channel $n' \to p'e'\nu'$.\footnote{Another possibility  
for solving the tension between the trap and beam lifetimes 
via  the neutron dark decay into mirror neutron $n \to n' \gamma(\gamma')$  was proposed in \cite{INT} 
(interestingly, it  can also destabilize the hydrogen atom via the decay  
${\rm H} \to n' \nu$ \cite{LHEP}). 
The same decay  
with $n'$ considered as an elementary fermion $\chi$ closely degenerate in mass with the neutron, 
was discussed also in Ref. \cite{Fornal}. 
However, $n\to n'$ decay solution is disfavored by the present experimental situation \cite{Abele}. 
As for $n-n'$ oscillation solution \cite{lifetime}, 
it is not excluded by now and can be tested in the experiments planned at the ORNL \cite{SNS}.  }
However, in the following we do not bias ourselves with this range of $\delta $ 
and consider it as a free parameter,   
large enough to satisfy the condition $\delta  \, t \gg 1$ in the 
oscillation experiments with the neutron flight time $t$. 

Without loss of generality, direction of the magnetic field can be taken  
as $z$-axis, i.e. $\vect{B}=B (0,0,1)$. Then $8\times 8$ Hamiltonian (\ref{H44}) 
can be split two $4\times 4$  matrices $H_+$ and $H_-$ 
describing separately the neutrons of two polarizations (spin projections $\pm1/2$ on $z$-axis): 
 \begin{equation}\label{H-mass}
 H_\pm =\PM{ - \delta  \mp \Omega   & 0 & \alp & \bet \\
0 &   - \delta \pm \Omega  & \bet  &  \alp \\
\alp &  \bet & 0  & 0  \\
\bet & \alp  & 0  &  0  }     
\end{equation}
where $\Omega= \vert \mu B \vert$.  
Now the probabilities of $n-n'$ and $n-\nbar'$ oscillations depend on the neutron polarization. 
In the case when $\vert \delta \pm \Omega \vert t \gg 1$, the time oscillations can be 
averaged and the mean probabilities read
\begin{eqnarray}
&& \ov{P}_{nn'}^\pm = \frac{2 \alp^2}{(\delta  \pm \Omega)^2 } , \quad 
\ov{P}_{n\nbar'}^\pm = \frac{2 \bet^2}{(\delta  \pm \Omega)^2 }  \label{nnpr} \\
&& \ov{P}_{\nbar n'}^\pm = \frac{2 \alp^2}{(\delta  \mp \Omega)^2 } , \quad 
\ov{P}_{\nbar\nbar'}^\pm = \frac{2 \bet^2}{(\delta  \mp \Omega)^2 }  \label{nprn}
\end{eqnarray}
Hence, the neutron disappearance probability $P_{nn'}^\pm + P_{n\nbar'}^\pm$ 
resonantly increases for $\Omega \approx \delta$ only for one spin state 
(namely for `$+$' polarization if $\delta <0$). 

As for the probability of induced $n-\nbar$ transition, it is polarization independent:  
 \begin{equation}\label{nnbar}
 P_{n\nbar}^\pm(t) = \frac{4\alp^2\bet^2}{(\delta^2 -\Omega^2)^2} + 
 \frac{4\alp^2\bet^2\delta^2 }{ (\delta^2 -\Omega^2)^2} \, \frac{\sin^2(\Omega t)}{\Omega^2}
  \end{equation}
Here  the first term is in fact the product $\ov{P}_{nn'}^\pm \ov{P}_{n\nbar'}^\mp$. 
However, for small magnetic field, $\Omega t \ll 1$, the second term is dominant  and we get 
$P_{n\nbar}(t) = (\dm^{\rm eff} t)^2$ which in fact mimics the contribution of direct $n-\nbar$ mixing 
(\ref{n-nbar}) with $\dm^{\rm eff}=2\alp\bet/\delta$. 
Hence, the experimental limit $\tau_{n\nbar} > 0.86 \times 10^8$\,s implies 
$\dm^{\rm eff}< 7.7 \times 10^{-24}$ eV and so 
\begin{equation}\label{dm-limit}
\alp\bet < (\delta/10^{-7}~{\rm eV}) \times 7.7 \cdot 10^{-31}~{\rm eV}^2 
\end{equation}  
We see from Eq. (\ref{nnbar}) that large magnetic field suppresses $n-\nbar$ oscillation. 
Only in quasi-free resonance regime $\vert \delta - \Omega\vert t < 1$ 
we get again $P_{n\nbar}(t) \sim (\dm^{\rm eff}\,  t)^2$.  
Thus, the presence of large mass splitting $\delta$ prevents the possibility 
of strong enhancement of $n-\nbar$ oscillation in constant magnetic field. 
However, larger resonance effect $P_{n\nbar} \simeq \alp^2\bet^2/(\delta-\Omega)^4$ 
can be obtained in the experiment in which magnetic field direction is changed 
to the opposite between the first and second parts of the neutron propagates vessel. 
Let us also remark that bound (\ref{dm-limit}) 
becomes invalid in the presence of reasonably large mirror magnetic field, 
or if there exist long range fifth forces mediated by 
hypothetical $\rB\!-\!\rL$ baryophotons which could induce level splitting up to $\sim 10^{-11}$~eV 
between $n$ and $\nbar$ states \cite{B-L,Babu:2016}.

\medskip 

\noindent
{\bf 5.} In the previous section we have shown that the strong effects of $n-\nbar$ conversion 
due to simultaneously present $n-n'$ and $n-\nbar'$ mixings are not excluded by 
existing experimental limits. 
Let us discuss now whether this possibility can be excluded by the nuclear stability bounds. 

As it is well known,  the direct $n-\nbar$ mixing leads to nuclear instability:   
the neutron inside the nuclei can annihilate with other nucleons 
into pions with total energy roughly equal to two nucleon masses. 
 The neutron-antineutron oscillations  in nuclei via  direct $n-\nbar$ mixing
is described by the Hamiltonian 
\begin{equation}\label{H22} 
H =\PM{ U   & \dm  \\
\dm  &  \ov{U}   }  , \quad U=-V, ~~ \ov{U} = -\bar{V}  - i W 
\end{equation}
where $V $ is the nucleon binding energy typically of few MeV,   
while the antineutron potential  $\ov{U}$ 
has both real and imaginary (absorptive) parts both being of the order of 100 MeV \cite{Gal}. 
The absorptive part $W$ is in fact related to the antineutron annihilation in nuclear medium. 

This Hamiltonian can be diagonalized by canonical transformation 
$S H S^{-1} = \widetilde H$ where $\widetilde H$ is diagonal and 
unimodular mixing matrix $S$ can be approximated as 
\begin{equation}\label{canonic} 
S =  \PM{S_{1n} & S_{1\nbar}  \\
S_{2n}  & S_{2\nbar}  } \approx 
\PM{1 & \dm/\Delta U \\
- \dm/\Delta U & 1 } 
\end{equation}  
with the precision up to quadratic terms in small ratio $\dm/\Delta U$,  
where $\Delta U = U- \ov{U}$. 

The neutron stationary state in nuclei can be viewed as a superposition 
$n_1 = S_{1n} n + S_{1 \nbar} {\nbar}$, and corresponding eigenvalue in  $\widetilde H$ 
also acquires a small imaginary part due to mixing.  
Thus $n_1$ contains the antineutron with the weight $S_{1\nbar}$ and 
so it can annihilate with other nucleons ($N=p,n$) producing pions.   
The rate  of this reaction can be readily estimated as
$\Gamma_{nN} = \vert S_{1\nbar}\vert^2 \Gamma_{\nbar}$, where 
$\Gamma_{\nbar N}= 2W$  is the antineutron annihilation width at nuclear densities. 
Thus, from (\ref{canonic}) we get: 
 \begin{equation}\label{nnbar-lifetime}
\Gamma_{nN} = \frac{\dm^2}{\vert \Delta U \vert^2}\Gamma_{\nbar N} 
= \frac{2 \dm^2 W}{(\ov{V}-V)^2 + W^2} 
\end{equation} 
Namely, the limit $ \dm^{-1} = \tau_{n\nbar} >  2.7 \times 10^8$\,s  
on $n-\nbar$ direct  mixing (\ref{dm}) can be obtained in this way 
from the experimental bound on $^{16}$O  stability 
$\Gamma_{nN} < (1.9 \times 10^{32}\,{\rm yr})^{-1}$ \cite{SK}. 
This limit is about 3 times stronger than 
the direct experimental limit $\tau_{n\nbar} > 0.86 \times 10^8$~s \cite{Baldo}. 

As it was observed in Ref. \cite{BB-nn'}, only $n-n'$ mixing cannot destabilize the nuclei by 
kinematical reasons (energy conservation). 
But in combination with $n-\nbar'$ mixing it in principle can,  as far as
$n-\nbar$ mixing  emerges at second order from $n-n'$ and $n-\nbar'$ mixings.

 Now the neutron oscillations in nuclear medium is described again by Hamiltonian (\ref{H44}) 
where we take the neutron and antineutron potentials $U$ and $\ov{U}$ as in (\ref{H22}).  
 Mirror states have vanishing 
potentials, $U', \ov{U}'=0$, and the magnetic contributions are negligible. 
Let us also set $\dm=0$ for excluding the effect of direct $n-\nbar$ mixing.  
Then mixing matrix (\ref{S}) can be directly computed:  
$S_{1n'} = -\alp/V$,  $S_{1\nbar'} = -\bet/V$,  and 
\begin{equation}\label{S-1nbar}
\vert S_{1\nbar}\vert  = \frac{2\alp\bet}{\vert V(\ov{V} -iW)\vert} 
\ll   \frac{\alp\bet}{V^2} = S_{1n'} S_{1\nbar'} 
\end{equation}     
Considering now that the neutron stationary state in nuclei  is the eigenstate 
$n_1=S_{1n} n + S_{1 \nbar} {\nbar}+ S_{1n'} n' + S_{1 \nbar'} {\nbar'}$,  
 It can annihilate with other ``spectator" nucleons ($N=p,n$) into pions,  
with the rate  $\Gamma_{nN} = \vert S_{1\nbar}\vert^2 \Gamma_{\nbar N}$. 
 In addition, two neutrons in the nucleus can also annihilate  
into (invisible) mirror pions, with the rate 
$\Gamma'_{nn}  \simeq \vert S_{1n'} S_{1\nbar'}\vert^2 \Gamma_{\nbar N}$. 
 For our benchmark $\tau_{nn'}\tau_{n\nbar'} \sim 100$~s$^2$, 
both these rates are below $(10^{60}~{\rm yr})^{-1}$. 
 For comparison,  Super-Kamiokande limit on $^{16}$O decays yields 
$\Gamma_{n N} < (1.9 \times 10^{32}~{\rm yr})^{-1}$ \cite{SK}   
 while  KamLAND limit on the pionless $^{12}$C decays  is  
$\Gamma'_{nn} < (1.4 \times 10^{30}~{\rm yr})^{-1}$   \cite{Kamyshkov1,Araki}.  
Thus, we can conclude that our scenario is perfectly safe against the nuclear stability bounds. 

Concluding this section, let us briefly discuss the implications for the neutron stars. 
In difference from nuclei, in this case $n-n'$ transitions are not forbidden by 
cinematic reasons, and they would convert a neutron star in a mixed star composed 
of both ordinary and mirror neutrons. 
However, as it was noticed in Ref. \cite{BB-nn'}, for $\tau_{nn'}\sim 1$~s 
the conversion time would be much larger than the age if the universe $t_U$. 
More detailed analysis brings to the conservative lower limit  $\tau_{nn'} > 10^{-2}\div 10^{-1}$~s or so 
 \cite{Massimo,Goldman}. 
 
 In the case when both $n-n'$ and $n-\nbar'$ mixings are present, with $\bet < \alp$, 
 then inside the neutron star  the neutron should undergo both $n-n'$ and $n-\nbar'$ transitions, 
with different rates respectively proportional to $\alp^2$ and $\bet^2$.
The produced $\nbar'$  would annihilate with $\nbar'$  into pions and this would 
could eventually bring to the evaporation of the neutron star. 
 However, for $\tau_{n\nbar'} > 1$~s the evaporation time would be two orders of magnitude 
 larger than $t_U$. In fact, the evaporation time comparable with $t_U$ can even have interesting 
 implications. In fact, when the mass of neutron star, due to this evaporation, 
 reduces to $0.1~M_\odot$ or so, it becomes unstable and should explode 
 producing the  hot and relatively dilute neutron rich medium 
 which can be associated to the kilonovae. 
 This phenomenon can be at the origin of r-processes 
 that enrich the universe by rare elements as gold etc.   
 
\medskip 
\noindent 
{\bf 6.} Concluding, we have discussed a fascinating possibility how the neutron 
could travel to parallel mirror world and then effectively return to our world as the antineutron. 
 The reason why this effect could skip  immediate detection in the experiments 
 can be related to the environmental factors, 
as the existence of the mirror magnetic fields at the Earth \cite{More}.  
This hypothesis may sound not so weird considering that there can exist 
dynamical interactions between ordinary and mirror particles mediated e.g. by the 
photon--mirror photon kinetic mixing $\frac{1}{2} \epsilon F^{\mu\nu} F'_{\mu\nu}$ \cite{Holdom} 
which effectively makes mirror particles electrically mini-charged. 
The cosmological bounds on the mixing parameter 
imply $\epsilon > 10^{-9}$ or so \cite{Ciarcelluti,Lepidi} 
while the direct experimental limit from the positronium decay reads
$\epsilon >5 \times  10^{-8}$ \cite{Crivelli}. 
The Rutherford-like interactions between free ordinary and mirror electrons 
mediated by this kinetic mixing  in rotating protogalaxies  can induce circular electric currents
which may be strong enough to create the observed magnetic fields on the galaxy scales 
with the help of moderate dynamo amplification \cite{BDT}. 
Such interactions can also have interesting implications for the direct detection of dark mirror 
matter \cite{DAMA2,DAMA1} which is expected be rather light as it should be dominated 
by the helium component \cite{BCV}. 
Ordinary and mirror particles can also interact  via  some common gauge bosons related e.g. to 
flavor symmetry \cite{PLB-su3,EPJ-Belfatto}.  

The sun and the Earth could capture some tiny amount of mirror matter via these interactions 
and this captured component  should be partially ionized  due to the high temperature at 
the centre of the Earth \cite{Anti-DM}. 
Then the drag of free mirror electrons by the Earth rotation can induce reasonably large 
circular mirror currents via the mechanism of Ref. \cite{BDT}. 
These circular currents can give rise to mirror magnetic field which can be 
further amplified by the dynamo mechanism. In this case, one can generically expect 
the mirror magnetic field at the Earth to vary in time. In particular, 
having a dipole character, it can change its direction with a relative periodicity of  
few years. 

The results of the neutron disappearance experiments 
\cite{Ban,Serebrov,Bodek,Serebrov2,Altarev,ILL,Abel} 
still allow $n-n'$ and/or $n-\nbar'$ oscillations with effective time 
$\tau = (\alp^2+\bet^2)^{-1/2}$ which can be as small as few seconds 
provided that mirror magnetic field $B'$ is larger than 0.5~G or so. 
Such a short oscillation time can have  interesting implications for the propagation 
of ultra-high energy  cosmic rays at cosmological distances \cite{UHECR,Askhat}. 
Interestingly, some experiments show deviations from null-hypothesis 
compatible with $\tau \simeq 20$~s provided that $B' \sim 0.1-0.3$ G \cite{Nesti,ILL}.  

A new UCN experiment planned at the PSI  can definitely test the existing anomalies 
with a sensitivity up to $\tau \sim 100$~s for $B' < 1$~G or so \cite{PSI}.  
Exclusion in this range would imply  $\tau_{nn'} \tau_{n\nbar'} > 10^{4\div5}$~s$^2$  
in which case Eq. (\ref{n-nbar-new}) still allows $P_{n\nbar} \sim 10^{-12}\div 10^{-14}$, 
which is anyway orders of magnitude larger than the maximal value $\sim 10^{-18}$ 
(\ref{n-nbar}) allowed in the case of direct $n-\nbar$ mixing (\ref{dm}).  
Once again, for amplifying the probability of $n-\nbar$ transition in the direct search experiments, 
the magnetic field $B$ should be properly scanned for finding the resonant value 
(unknown mirror field $B'$)  
instead of suppressing it as in the classic experiment  \cite{Baldo}.

For testing this possibility,  experiments on $n-\nbar$ search should be 
performed by scanning different values of magnetic field. 
instead of suppressing it  as in classic experiment \cite{Baldo}. 
In particular, this can be done in $n\to n' \to n$ regeneration experiments 
planned  at the ORNL \cite{ORNL},  
by modifying it for searching $n\to n'/\nbar' \to \nbar$ process  
by replacing the neutron detector by the antineutron one. 

The discussed possibility of fast $n-n'$ and $n-\nbar'$ oscillations points towards 
the scale $M\sim 10$~TeV
for underlying new physics responsible for operators (\ref{B+B}) and (\ref{B-B})  
which can be reachable in future accelerators. 
In addition, it could suggest a low scale baryogenesis scenario  generating 
baryon asymmetries in both sectors, 
Namely, baryon asymmetries can be produced, from the couplings (\ref{Yuk}) and (\ref{Yuk-2})
 via out-of equilibrium $\rB$, $\rB'$ and CP violating scattering processes 
$dS \to d'S'$ and $dS \to d^{\prime c} \ov{S}'$ involving color scalars $S$ and $S'$ 
of two sectors, along  the lines of the co-leptogenesis mechanism 
related to ordinary and mirror neutrino mixings discussed in Refs. \cite{BB-PRL}.

\begin{acknowledgments}
\noindent
I thank Yuri Kamyshkov for valuable discussions. 
 The work was supported in part by Ministero dell'Istruzione, Universit\`a e della Ricerca (MIUR) 
 under the program PRIN 2017, Grant 2017X7X85K 
``The dark universe: synergic multimessenger approach",  
and in part by the Shota Rustaveli National Science Foundation (SRNSF)   
Grant DI-18-335/New Theoretical Models for Dark Matter Exploration.
\end{acknowledgments}

\end{document}